\documentclass[
aps,
pra,
notitlepage,
citeautoscript, superscriptaddress,
reprint
]{revtex4-2}
\usepackage[dvipsnames,usenames]{xcolor}
\usepackage{graphicx}
\usepackage{graphicx}
\usepackage{dcolumn}% Align table columns on decimal point
\usepackage{bm}% bold math
\usepackage{mathrsfs,natbib}
\usepackage[title]{appendix}
\usepackage{graphicx,epsf,amssymb,amsbsy,amsfonts,amssymb,amsmath}
\usepackage[colorlinks=true]{hyperref}
\hypersetup{
    unicode=false,          % non-Latin characters
    pdftoolbar=true,        % show Acrobat
    pdfmenubar=true,        % show Acrobat
    pdffitwindow=false,     % window fit to page when opened
    pdfstartview={FitH},    % fits the width of the page to the window
    pdftitle={},    % title
    pdfauthor={},     % author
    pdfsubject={},   % subject of the document
    pdfcreator={},   % creator of the document
    pdfproducer={}, % producer of the document
    pdfkeywords={} {} {}, % list of keywords
    pdfnewwindow=true,      % links in new window
    colorlinks=true,       % false: boxed links; true: colored links
    linkcolor=magenta, %red,          % color of internal links (change box color with linkbordercolor)
    citecolor=blue,        % color of links to bibliography
    filecolor=magenta,      % color of file links
    urlcolor=blue           % color of external links
}
\usepackage{slashed}
\usepackage{comment}
\usepackage{color,soul}

\newcommand{\be}{\begin{equation}}
\newcommand{\ee}{\end{equation}}
\newcommand\beq{\begin{eqnarray}}
\newcommand\eeq{\end{eqnarray}}

\newcommand\ket[1]{| #1 \rangle}

\newcommand{\mybar}[1]

\begin{document}

\title{Floquet insulators and lattice fermions}% Force line breaks with \\
\author{Thomas Iadecola}
\email{iadecola@iastate.edu}
\affiliation{Department of Physics and Astronomy, Iowa State University, Ames, Iowa 50011, USA}%
\affiliation{Ames National Laboratory, Ames, Iowa 50011, USA}%
\author{Srimoyee Sen}%
\email{srimoyee08@gmail.com}
\affiliation{Department of Physics and Astronomy, Iowa State University, Ames, Iowa 50011, USA}%
\author{Lars Sivertsen}%
\email{lars@iastate.edu }
\affiliation{Department of Physics and Astronomy, Iowa State University, Ames, Iowa 50011, USA}%

\date{\today}

\begin{abstract}
Floquet insulators are periodically driven quantum systems that can host novel topological phases as a function of the drive parameters. These new phases exhibit features reminiscent of fermion doubling in discrete-time lattice fermion theories. We make this suggestion concrete by mapping the spectrum of a noninteracting (1+1)D Floquet insulator for certain drive parameters onto that of a discrete-time lattice fermion theory with a time-independent Hamiltonian. The resulting Hamiltonian is distinct from the Floquet Hamiltonian that generates stroboscopic dynamics. It can take the form of a discrete-time Su-Schrieffer-Heeger model with half the number of spatial sites of the original model, or of a (1+1)D Wilson-Dirac theory with one quarter of the spatial sites.  
\end{abstract}

%\pacs{Valid PACS appear here}

\maketitle

Zero-energy fermionic modes, or zero modes, are among the earliest manifestations of topology in quantum many-body theory. In systems with a gapped bulk, they appear localized at topological defects like solitons and vortices, or as dispersive modes at boundaries between phases with different bulk topological invariants~\cite{TeoKane}. They can carry fractional charges under global symmetries and manifest fractional or non-Abelian statistics when braided~\cite{JackiwRebbi,JackiwRossi,HCM,MooreRead,ReadGreen,Ivanov}, and their existence enforces degeneracies in the many-body spectrum~\cite{Fendley12,Fendley16}.
% A rich theory of these modes and their descendants has emerged over the last five decades.

Quantum systems driven periodically in time, known as Floquet systems, furnish intrinsically nonequilibrium generalizations of topological and conventionally ordered phases~\cite{CayssolReview,RudnerReview,SachaReview,ElseReview,KhemaniReview,MonroeDTC,LukinDTC,MonroePDTC,GoogleDTC,DumitrescuEDSPT,DengFSPT,McIver19}. Despite lacking energy conservation, they retain a notion of eigenstates and and eigenvalues when observed at ``stroboscopic times" that are integer multiples of the driving period $T$. Instead of an energy spectrum which can in principle be unbounded in the thermodynamic limit, stroboscopic dynamics and eigenstates are characterized by a bounded spectrum of quasienergies $-\pi/T\leq \epsilon <\pi/T$ that are only conserved modulo $\frac{2\pi}{T}$. The periodic nature of quasienergy furnishes a generalization of zero modes. For example, fermionic Floquet systems known as Floquet insulators can exhibit localized ``$\pi$ modes" whose existence implies a ``$\pi$ pairing" between many body states at quasienergy $\epsilon$ and $\epsilon+\frac{\pi}{T}$~\cite{Thakurathi,Rudner13,Khemani,ElseTC,ElseFSPT,vonKeyserlingk}. 
% $\pi$ modes and related phenomena such as discrete time crystals~\cite{SachaReview,ElseReview,KhemaniReview} are viewed as intrinsically nonequilibrium and unique to the driven setting.

\begin{figure}
\includegraphics[width = .9\columnwidth]{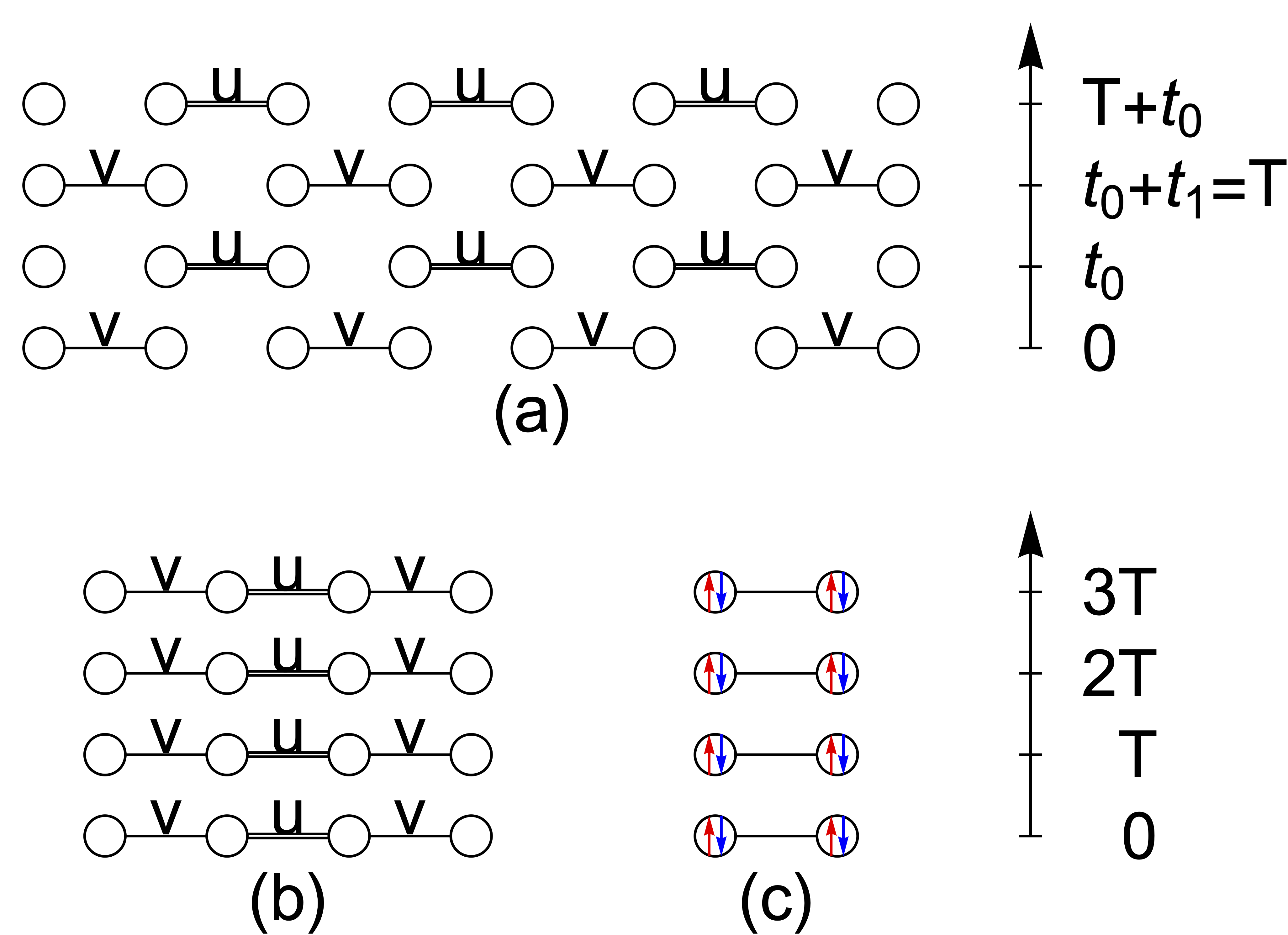}
\caption{Schematic of the mapping between stroboscopic Floquet dynamics and time-independent lattice Hamiltonians. The spectrum of the stroboscopic dynamics corresponding to Eq.~\eqref{uF}, shown in (a), can be mapped onto that of a discrete-time Su-Schrieffer-Heeger (b) or Wilson-Dirac (c) theory. The number of spatial sites in (b) is halved with respect to (a) to accommodate the additional degrees of freedom due to fermion doubling. The number of sites are further halved in (c) to accommodate the spinful nature of the Wilson-Dirac fermions (depicted via arrows inside the lattice sites).}
\label{illustration}
\end{figure}

Another setting in which $\pi$ modes arise is spacetime lattice regularizations of fermionic quantum field theories. In particular, discretizing the Dirac operator on a spacetime lattice leads to the so-called fermion doubling problem, where the total number of fermionic degrees of freedom increases by a factor of $2^D$, where $D$ is the spacetime dimension \cite{Nielsen:1980rz, Nielsen:1981xu}. The extra modes, known as ``doublers," are undesirable and several methods, including Wilson, Kogut-Susskind, and domain-wall fermions \cite{Kaplan:1992bt, Shamir:1993zy, PhysRevD.25.2649, Neuberger:1997fp, Neuberger:1998wv, PhysRevD.11.395}, are commonly used to dispense with them. Nevertheless, it is natural to ask whether the doubler modes associated with zero modes, which occur at frequency $\frac{\pi}{\tau}$ where $\tau$ is the temporal lattice spacing, are related to the $\pi$ modes that can appear in Floquet insulators.

In this paper, we answer this question affirmatively for a particular (1+1)D Floquet-insulator model. Specifically, we show that the quasienergy spectrum of a continuous-time Floquet model of spinless complex fermions can be mapped directly onto the spectrum of discrete-time lattice fermions with a time-independent Hamiltonian. The lattice spacing of this discrete time theory equals the driving time period $T$ of the Floquet system. The mapping is made easier by the observation that the Dirac equation takes the form of a Schroedinger equation for the corresponding fermion Hamiltonian. The lattice fermion model can take the form of a Su-Schrieffer-Heeger (SSH) model with half the spatial lattice sites of the Floquet model, or of a Wilson-Dirac model with one quarter of the sites. Furthermore, for appropriately chosen solutions of the mapping between spectra, the phase diagrams of the Floquet and lattice models match. In particular, the topological phase of the Floquet model, which exhibits localized zero and $\pi$ modes with open boundary conditions, coincides with the topological phase of the related lattice Hamiltonians, where $\pi$ modes appear as doublers of zero modes. 
% For some previous studies analyzing the ties of lattice fermions and  Floquet systems see \cite{PhysRevLett.121.196401, demarco2018single}.
While previous studies~\cite{PhysRevLett.121.196401, demarco2018single} have used Floquet systems to provide new perspectives on lattice fermion doubling, here we focus on exposing a direct correspondence between the two.
% In the remainder of the paper, we will briefly review lattice fermion doubling, define the Floquet model we consider, and then explain the various mappings.

\textbf{Lattice fermion doubling.}---A continuous-time free-fermion theory in Minkowski space with Hamiltonian $H$ has the fermion/Dirac operator $\gamma_0(i\partial_t-H)$ from which we can extract the Schr\"odinger operator $i\partial_t-H$. The eigenvalues of this operator are $\pm\sqrt{p_0^2-E^2}$, where $p_0$ is the Fourier variable conjugate to time $t$ and $E$ are the eigenvalues of $H$~\footnote{Unless indicated otherwise, we restrict our focus in this work to single-particle spectral properties rather than many-body ones.}. The zeroes of these eigenvalues at $p_0=E$ correspond to poles of the fermion propagator $(i\partial_t-H)^{-1}\gamma_0^{-1}$. Discretizing time in this theory leads to fermion doubling \cite{Nielsen:1980rz, Nielsen:1981xu}. To see this, let $\tau$ be the lattice spacing in the time direction. The poles of the discrete-time theory now satisfy $\frac{1}{\tau}\sin(p_0\tau) = E$, which leads to two solutions:
\begin{align}
\label{poles}
 p_0=\frac{1}{\tau}\sin^{-1}(E\tau)\indent \text{and} \indent p_0=\frac{\pi}{\tau}-\frac{1}{\tau}\sin^{-1}(E\tau)   
\end{align}
Thus, for every pole of the continuous-time propagator there are two poles of the discrete-time propagator \cite{GuyMoore,Ambjorn:1990pu,Aarts:1998td, Mou:2013kca}. In particular, for a zero mode of the continuous-time theory, the discrete-time theory has modes at $p_0=0$ and $p_0=\frac{\pi}{\tau}$; the new $\pi$ mode arises purely due to time discretization.

A corresponding phenomenon is also observed in Euclidean-time lattice field theory \cite{Karsten:1981gd}~\footnote{Typical lattice Monte Carlo studies of quantum field theories are performed in Euclidean space-time. For real-time formulations path integrals for quantum field theories and quantum mechanical systems see \cite{ Kanwar:2021tkd, Alexandru:2016gsd, Lawrence:2021izu, Tanizaki:2014xba, Alexandru:2017lqr, Mou:2019tck}}. There, the Schr\"odinger operator in continuous time has eigenvalues $\pm\sqrt{p_0^2+E^2}$, which, unlike the Minkowski theory, vanishes only at $p_0=E=0$. In this case, upon discretizing time, fermion doubling manifests itself as a degeneracy of Schr\"odinger eigenvalues. For instance, any eigenstate $\ket{p_0,E}$ with eigenvalue $\frac{1}{\tau}\sqrt{\sin^2(p_0\tau)+E^2}$ has a partner $\ket{\frac{\pi}{\tau}-p_0,E}$ with the same eigenvalue. In congruence with the Minkowski-space analysis, if the continuous-time fermion propagator has a pole at $p_0=E=0$, then there is another pole at $p_0=\pi/\tau$.
More interestingly, the states $\ket{E,E}$, which correspond to the energy eigenstates of $H$, have the same spatial profiles in the position eigenbasis as $\ket{\frac{\pi}{\tau}-E,E}$. Therefore, if $H$ has a spatially localized zero mode, the corresponding discrete-time Schr\"odinger equation has two solutions with the same spatial profile.  

In this paper, we will consider real-time (i.e., Minkowski) theories as opposed to imaginary-time (Euclidean) ones, as the comparison between the Floquet and lattice spectra is more natural for the former. However, there is no fundamental obstruction to performing a similar analysis for Euclidean theories, and such an approach may be desirable when generalizing beyond the case of noninteracting fermions, which is our focus here.
% Indeed, interacting fermion field theories are typically analyzed in Euclidean spacetime.
%However, real-time lattice field theory approaches are sought after for real-time dynamics of heavy ion collisions, early-universe physics, and other applications. 
We will briefly mention how the comparison between Floquet systems and lattice fermions in Minkowski spacetime can be adapted for Euclidean spacetime lattices.

\textbf{Floquet insulator model.}---As a simple example of a Floquet insulator, we consider a continuous-time (1+1)D model on a spatial lattice of $2N$ sites defined by the evolution operator
\begin{align}
U(t)&=\begin{cases}e^{-i H_0 t} & \text{for } 0<t<t_0\\
e^{-i H_1(t-t_0)}e^{-i H_0 t_0} & \text{for } t_0\leq t<t_0+t_1
\end{cases},
\label{u}
\end{align}
with
\begin{align}
\begin{split}
H_0&=2\sum^{N-1}_{j=0}(a_{2j}^{\dagger}a_{2j+1}+\text{H.c.})\\
H_1&=2\sum^{N-1}_{j=0}
(a_{2j+1}^{\dagger}a_{2j+2}+\text{H.c.}),
\label{h}
\end{split}
\end{align}
where $a_i$ is a fermion annihilation operator on site $i=0,\dots,2N-1$. Unless otherwise specified, we consider periodic boundary conditions (PBC) such that $a_{2N}\equiv a_0$.
The Hamiltonians in Eq.~\eqref{h} are the ``trivial" and ``topological" parts of the (static) SSH model~\cite{SSH,JackiwRebbi},
\begin{align}
\label{hssh}
    H_{\rm SSH}=\frac{u}{2}H_1+\frac{v}{2}H_0.
\end{align}
The energy spectrum of this model with PBC is given by \begin{align}
\label{essh}
    E_{\rm SSH}(k) = \pm\sqrt{u^2+v^2+2uv\cos(2k)},
\end{align}
where $0\leq k < \pi$ is the crystal momentum and we have set the spatial lattice spacing to 1. The SSH model has a symmetry-protected topological (SPT) and a trivial phase that have identical bulk spectra. The two gapped phases can be distinguished by their energy spectra in open boundary conditions (OBC): in the SPT phase there are two spatially localized zero modes pinned to the middle of the energy gap, one at each end of the chain, while in the trivial phase there are none. Expanding the model around $k=\frac{\pi}{2}$ yields a Dirac fermion theory with a mass proportional to $u-v$; the transition between the SPT and trivial phases occurs at the massless point $u=v$. $H_0$ and $H_1$ can be viewed as representatives of the trivial and SPT phases, respectively.

\begin{figure}[t]
\includegraphics[width=0.5\columnwidth]{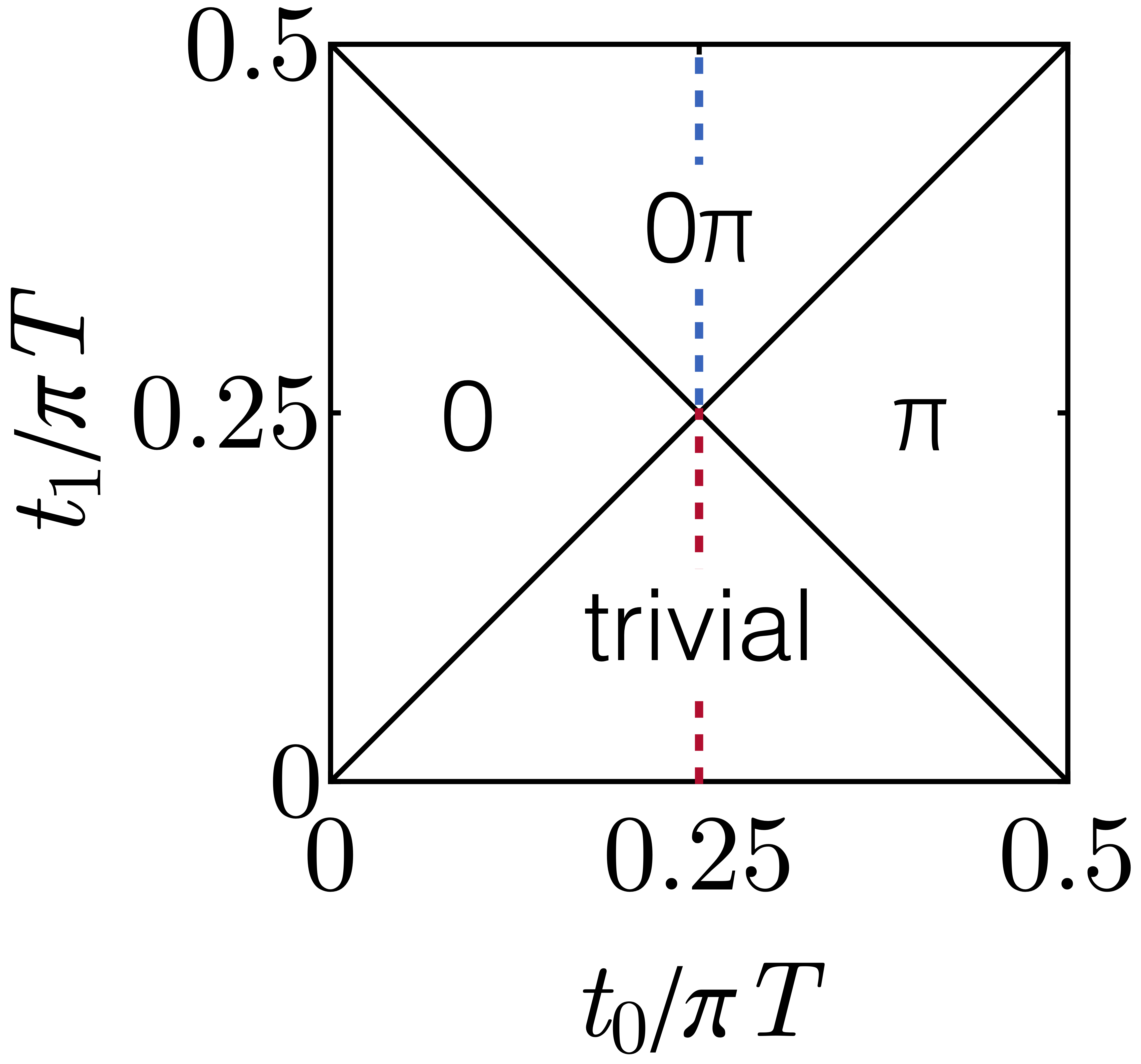}
\caption{Phase diagram of the Floquet model~\eqref{uF}. The phases are labeled by the presence or absence of zero and $\pi$ modes localized to boundaries. This work focuses on the vertical dashed line at $\frac{t_0}{T}=\frac{\pi}{4}$, which passes through the trivial and $0\pi$ phases.
}
\label{fig:phase diagram}
\end{figure}

The Floquet model~\eqref{u} is closely related to a simple (1+1)D model discussed in Refs.~\cite{vonKeyserlingk,Khemani,Potter}, with the small difference that Eq.~\eqref{u} is formulated in terms of complex rather than Majorana fermion operators~\cite{SM}. The analysis of both models is nearly identical, and they have the same phase diagram. Since energy is no longer conserved in the time-dependent model \eqref{u}, phases are classified according to spectral properties of the Floquet operator
\begin{align}
\label{uF}
U_{\rm F} = U(T) = e^{-iH_1t_1}e^{-iH_0t_0} \equiv e^{-iH_{\rm F}T},
\end{align}
where $T=t_0+t_1$ is the driving period and $H_{\rm F} = \frac{i}{T}\ln U_{\rm F}$
is known as the Floquet or stroboscopic Hamiltonian. The quasienergies $-\pi/T\leq\epsilon<\pi/T$ are the eigenvalues of $H_{\rm F}$. They can be obtained analytically because the model \eqref{uF} is quadratic~\cite{SM}. There are four phases, which we label trivial, $0$, $\pi$, and $0\pi$ (see Fig.~\ref{fig:phase diagram}). The $0$ and $\pi$ phases have localized boundary modes with $\epsilon=0$ and $\frac{\pi}{T}$, respectively; the $0\pi$ phase has both zero and $\pi$ modes, while the trivial phase has neither. The phase boundaries in Fig.~\ref{fig:phase diagram} are the lines in the $t_0$-$t_1$ plane where the quasienergy gap closes. The trivial and $0$ phases are adiabatically connected to the trivial and topological phases of the SSH model, respectively. The phases with $\pi$ modes are ``intrinsically Floquet" phases in the sense that they cannot arise in the absence of the drive.

\begin{figure}[t]
\includegraphics[width=\columnwidth]{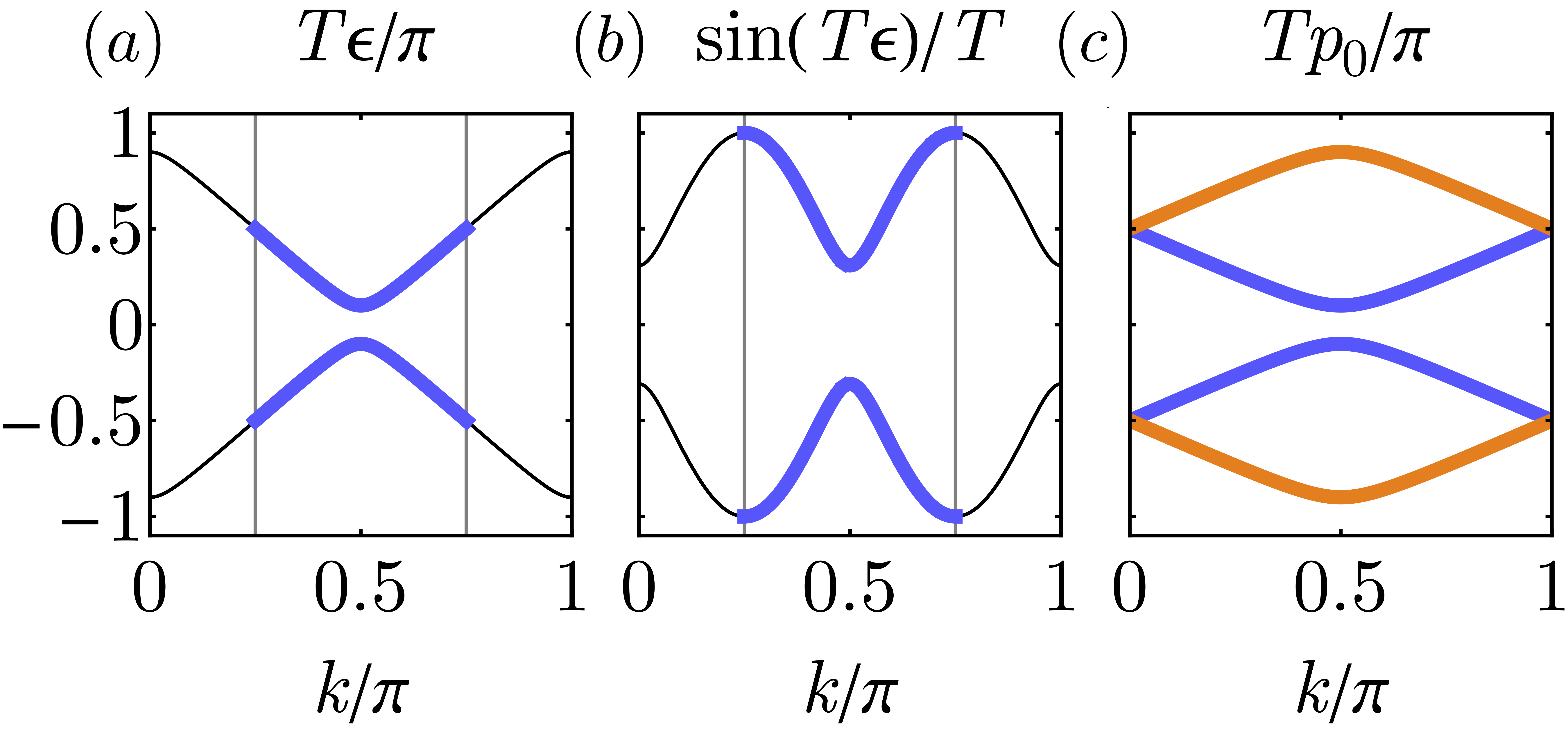}
\caption{Spectrum of the Floquet model and its mapping onto the lattice spectrum. (a) Quasienergy spectrum of Eq.~\eqref{uF} along the line $\frac{t_0}{T}=\frac{\pi}{4}$ [see Eq.~\eqref{eq:epsilon}]. The quasienergy values highlighted in blue are those corresponding to crystal momenta $k$ in the interval $[\frac{\pi}{4},\frac{3\pi}{4})$ (grey vertical lines), denoted $\tilde\epsilon$ in the text. (b) Sine-transformed quasienergy spectrum entering Eq.~\eqref{E}. (c) Pole positions $p_0$ with momenta assigned according to the SSH mapping. The orange bands denote doublers, which correspond to the quasienergies $\frac{\pi}{T}-\tilde\epsilon$.
}
\label{fig:epsilon}
\end{figure}

\textbf{Floquet to lattice mapping.}---It is natural to speculate that the $\pi$ modes in the Floquet system can be viewed as doubler modes of an appropriate discrete-time lattice fermion theory with time lattice constant $\tau=T$. To confirm this, we will identify a discrete-time theory with Hamiltonian $H$ and spectrum $E$ such that the poles in Eq.~\eqref{poles} are in one-to-one correspondence with quasienergy eigenvalues $\epsilon$. Note that, in order for this identification to be possible, it is necessary that the (single-particle) spectrum of $H_{\rm F}$ exhibit a $\pi$-pairing such that for every quasienergy $\epsilon$ there is a partner at $\frac{\pi}{T}-\epsilon$. For the model \eqref{uF}, this feature arises along the line $\frac{t_0}{T}=\frac{\pi}{4}$, which connects the trivial and $0\pi$ phases. 
On this line, the quasienergy spectrum of $H_{\rm F}$ with PBC takes the form~\cite{SM}: 
\begin{align}
\label{eq:epsilon}
\epsilon(k) = \pm\cos^{-1}[-\cos(2\eta)\cos(2k)],
\end{align}
where $\eta = \frac{t_1}{T}-\frac{\pi}{4}$ measures the distance from the gap closure at $\frac{t_1}{T}=\frac{\pi}{4}$.
To enable a one-to-one mapping onto the poles \eqref{poles}, we will partition the quasienergy spectrum $\epsilon$ into two subsets containing quasienergies $\tilde\epsilon$ and $\frac{\pi}{T}-\tilde\epsilon$. We will then seek a model $H$ whose energy spectrum $E$ satisfies
\begin{align}
\label{E}
E=\frac{1}{T}\sin(T\tilde\epsilon),
\end{align}
which stems from identifying $p_0=\tilde\epsilon$ in Eq.~\eqref{poles}. A suitable choice of $\tilde\epsilon$ consists of the values $\epsilon(k)$ for $\frac{\pi}{4}\leq k < \frac{3\pi}{4}$ (see Fig.~\ref{fig:epsilon}).
The corresponding construction in Euclidean spacetime aims to identify the eigenvalues of the states $\ket{E, E}$ and
$\ket{\frac{\pi}{T} -E, E}$ with $\frac{1}{T}\sqrt{2}\sin (T\tilde\epsilon)$.

\textbf{SSH mapping.}---We now seek a Hamiltonian whose spectrum satisfies Eq.~\eqref{E}. Since $U_{\rm F}$ is built using the SSH-type Hamiltonians \eqref{h}, we first consider an SSH Hamiltonian with dispersion given by Eq.~\eqref{essh}. In this case, Eq.~\eqref{E} can be solved by
\begin{align}
\label{esshp}
    E_{\rm SSH}(k') = \frac{1}{T}\sin\left[T\epsilon\left(\frac{k'}{2}+\frac{\pi}{4}\right)\right],
\end{align}
where $0\leq k'< \pi$, where
\begin{align}
\label{uv}
 u=\frac{1\pm\sin(2\eta)}{2T}\indent\text{and}\indent v = \frac{1}{T}-u.
\end{align}
(Note that we have assumed $u,v\geq0$ for simplicity.)
% \textcolor{red}{
% There is no a-priori reason to expect that Eq. \eqref{esshp} to hold for any choice of $u$ and $v$. In fact such a solution does not hold if we use the Floquet eigenvalues themselves instead of taking their sine, i.e. replace 
% \begin{align}
% \label{esshp2}
%     \frac{1}{T}\sin\left[T\epsilon\left(\frac{k'}{2}+\frac{\pi}{4}\right)\right] \rightarrow \left[T\epsilon\left(\frac{k'}{2}+\frac{\pi}{4}\right)\right],
% \end{align}
% on the RHS of Eq. \eqref{esshp}
% where $0\leq k'< \pi$. Remarkably, the sinusoidal of the Floquet eigenvalues can be mapped to an SSH Hamiltonian as evidenced by Eq. \eqref{esshp}. 
% } 
The two possible assignments of $u$ satisfying Eq.~\eqref{esshp} have no impact on the bulk energy spectrum, but can nevertheless be physically distinguished by solving the SSH model with OBC.
The SSH model~\eqref{hssh} is in the SPT phase when $u>v$ and the trivial phase when $u<v$. Since the topological $0\pi$ phase of $H_{\rm F}$ occurs when $\eta>0$, we must pick the branch of Eq.~\eqref{uv} such that $u>v$ when $\eta>0$; this is accomplished by picking the ``$+$" branch. Let $\tilde H_{\rm SSH}$ denote the resulting time-independent SSH Hamiltonian.

By construction, the doubled spectrum \eqref{poles} of $\tilde H_{\rm SSH}$ when defined on a discrete-time lattice with spacing $\tau=T$ [Fig.~\ref{fig:epsilon}(c)] matches the quasienergy spectrum of the Floquet Hamiltonian $H_{\rm F}$ [Fig.~\ref{fig:epsilon}(a)]. Thus, $\tilde H_{\rm SSH}$ cannot be defined on the same spatial lattice as $H_{\rm F}$; rather, if the original Floquet model is defined on a lattice of $2N$ sites, $\tilde H_{\rm SSH}$ must be defined on $N$ sites. To see this, recall that Eq.~\eqref{esshp} maps the interval $\frac{\pi}{4}\leq k< \frac{3\pi}{4}$ to the interval $0\leq k'< \pi$. The $k$ interval contains half of the $N$ allowed crystal momentum values for $H_{\rm F}$, so $k'$ can only take $\frac{N}{2}$ values (note that this requires $N$ to be even). This corresponds to $N$ sites because the SSH model has a two-site unit cell. Thus, $\tilde H_{\rm SSH}$ should not be interpreted as a mere rewriting of $H_{\rm F}$. Instead, the models are related by the nontrivial procedure of fermion doubling.

\textbf{Wilson-Dirac mapping.}---To make more direct contact with lattice field theory, we now show that Eq.~\eqref{E} can also be satisfied by the spectrum of a Wilson-Dirac (WD) Hamiltonian. The corresponding solution is inspired by the observation that the spectrum of a WD Hamiltonian with PBC can be mapped onto that of an SSH Hamiltonian with PBC. The WD Hamiltonian on a 1D spatial lattice with $N$ sites can be written (here working in OBC for simplicity)
\begin{align}
\label{hwd}
    H_{\rm WD}=\!\!\sum^{N-1}_{x,x'=0}\!\bar\psi_x\!\left[R\, \gamma_1\, (-i\nabla_{x,x'})-\frac{R}{2}\, \nabla^2_{x,x'}+m\, \delta_{x,x'}\right]\!\psi_{x'},
\end{align}
where $\nabla_{x,x'}=(\delta_{x',x+1}-\delta_{x',x-1})/2$ is a symmetric spatial derivative, $\nabla^2_{x,x'}=\delta_{x'-1,x+1}+\delta_{x'+1,x-1}-2$ is the second derivative, $\psi_x$ is a two-component Dirac spinor with associated gamma matrices $\gamma_0,\gamma_1$, and $\bar\psi_x=\psi^\dagger_x\gamma_0$. The energy spectrum of this model with PBC is given by
\begin{align}
E_{\rm WD}(p)=\pm\sqrt{R^2 \sin^2 p+[m+R(1-\cos p)]^2},
\end{align}
where $-\pi\leq p<\pi$. Solving Eq.~\eqref{E} yields
\begin{align}
\label{ewdp}
    E_{\rm WD}(p')=\frac{1}{T}\sin\left[T\epsilon\left(\frac{p'}{4}+\frac{\pi}{2}\right)\right],
\end{align}
where $-\pi\leq p'<\pi$, provided that
\begin{align}
m=\pm\frac{\sin(2\eta)}{T}\indent \text{and}\indent R=\frac{1}{2T}-\frac{m}{2},
\end{align}
where we have assumed $R\geq0$ for simplicity.
Following the SSH case, we now ask which of these branches corresponds to a topological phase when $\eta>0$. With OBC, the WD Hamiltonian~\eqref{hwd} exhibits localized edge modes when $m<0$~\cite{SM}. Demanding that this condition coincide with positive $\eta$ selects the ``$-$" branch. We denote the resulting time-independent Wilson-Dirac Hamiltonian by $\tilde H_{\rm WD}$. 

We have thus found a second time-independent Hamiltonian, $\tilde H_{\rm WD}$, whose doubled spectrum when defined on a discrete-time lattice coincides with the quasienergy spectrum of $H_F$. We note here that $\tilde H_{\rm WD}$ must be defined on a lattice with $N/2$ sites (i.e., one quarter of the sites in the original Floquet model). Like the SSH case, the crystal momentum interval $0\leq k< \frac{\pi}{2}$ corresponding to $\tilde\epsilon$ contains $\frac{N}{2}$ points; thus, so does the new crystal momentum interval $-\pi\leq p'< \pi$. However, unlike the SSH model, the WD model has a one-site unit cell, so there is one site per crystal momentum value.

\begin{figure}[t]
\includegraphics[width=\columnwidth]{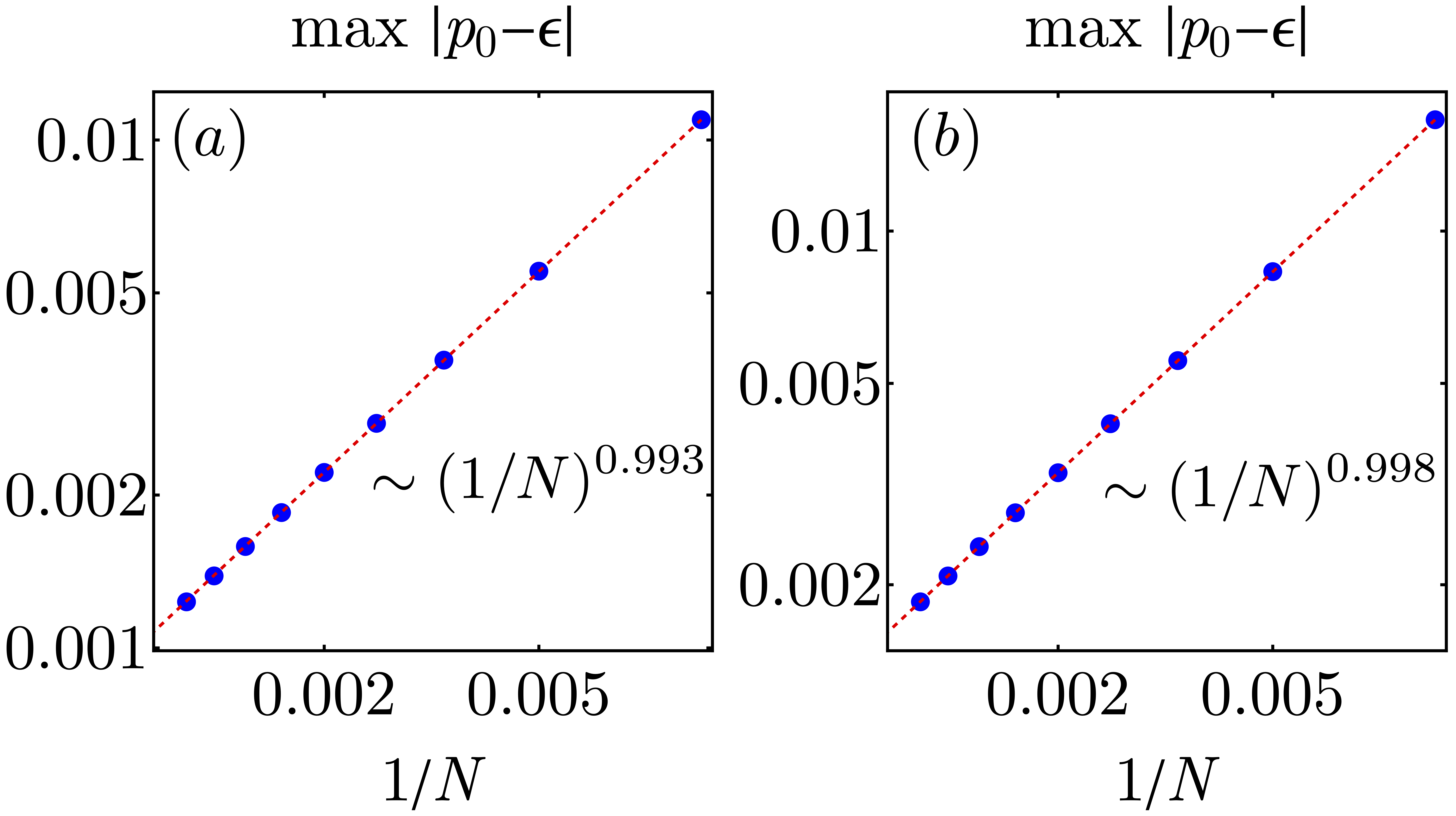}
\caption{
Finite-size scaling of the maximum difference between discrete-time spectra $p_0$ and quasienergy spectra $\epsilon$ (a) for OBC and (b) for OBC with a domain wall where $\eta$ switches sign. In both panels we compare to the SSH spectrum and fix $\eta=\pi/8$, considering system sizes $N=100,200,\dots,900$. Dashed lines indicate best fits to power laws in $1/N$ that are consistent with the expected scaling.
}
\label{fig:finite-size}
\end{figure}

\textbf{Discussion.}---The equalities \eqref{esshp} and \eqref{ewdp} relate the discrete-time spectra of static SSH and WD models to the quasienergy spectrum of Eq.~\eqref{uF} with PBC. In each case, the ``correct" branch of the respective model parameter solutions is chosen by matching the presence of zero modes as a function of $\eta$. However, these zero modes are properties of the OBC spectrum. Given that the PBC spectra match by construction, we expect that the OBC spectra match up to $\sim 1/N$ corrections. In Fig.~\ref{fig:finite-size}(a) we demonstrate this numerically for the SSH mapping by plotting the system size dependence of the maximum difference between the discrete-time frequency spectrum $p_0$ and the quasienergy spectrum $\epsilon$, where each spectrum is constructed as an ordered list. 

Another test of equivalence is to compare spectra in the presence of a domain wall in $\eta$. For the SSH model, this corresponds to a standard domain wall in the mass profile~\cite{JackiwRebbi}, while for the WD model it amounts to a domain wall in both the mass and the Wilson parameter $R$. For the Floquet model, a domain wall in $\eta$ entails an abrupt change in the coefficient of $H_1$ from $2$ to $2(\frac{\pi}{4}-\eta)/(\frac{\pi}{4}+\eta)$ at some point in space. In Fig.~\ref{fig:finite-size}(b) we examine the finite-size dependence of the difference between the discrete-time SSH and quasienergy spectra in the presence of a domain wall in the middle of the chain, again finding the expected $\sim 1/N$ scaling. These results build confidence that the mapping we develop defines a notion of equivalence in the thermodynamic limit between models irrespective of boundary conditions or the presence of topological defects.

\textbf{Outlook.}---We have demonstrated spectral equivalence between a simple Floquet insulator model in (1+1) dimensions and two canonical fermion models (SSH and WD) defined on a two-dimensional spacetime lattice. The mapping between spectra relies on a notion of $\pi$ pairing in the single-particle spectrum, which occurs along the line $\frac{t_0}{T}=\frac{\pi}{4}$. This pairing allows us to discard half of the degrees of freedom in the Floquet model to define a related static model, which recovers the discarded eigenvalues via fermion doubling after time discretization. One interesting question for future work is whether the equivalence can be extended off the line $\frac{t_0}{T}=\frac{\pi}{4}$~\cite{OngoingWork}. We note that $\pi$ pairing also occurs along the line $\frac{t_1}{T}=\frac{\pi}{4}$ for PBC but is lost for OBC, as evidenced by the separate existence of $0$ and $\pi$ modes with OBC. A separate direction is to consider extensions of these mappings to (2+1)D systems, where new phases like anomalous Floquet topological insulators emerge~\cite{Rudner13,Po}. In any dimension, it is also worth considering whether a similar equivalence between bulk topological invariants in Floquet systems~\cite{Kitagawa,Rudner13,Nathan,Carpentier,Fruchart} and lattice field theories can emerge \cite{Golterman:1992ub}. Finally, it will be interesting to consider whether an equivalence to lattice field theory can be formulated for interacting Floquet systems, which experience unbounded heating that requires the introduction of prethermalization or localization physics to enable long-lived phenomena~\cite{Ponte,D'Alessio,Khemani,Else17}.

\begin{acknowledgments}
TI acknowledges support from the National Science Foundation under grant DMR-2143635. SS and LS acknowledge support from the U.S.~Department of Energy,
Nuclear Physics Quantum Horizons program through the Early Career Award DE-SC0021892.
\end{acknowledgments}

\bibliography{floquet.bib}

\end{document}

\title{Supplemental materials for ``Floquet insulators and lattice fermions"}
\author{Thomas Iadecola}
\email{iadecola@iastate.edu}
\affiliation{Department of Physics and Astronomy, Iowa State University, Ames, Iowa 50011, USA}%
\affiliation{Ames National Laboratory, Ames, Iowa 50011, USA}%
\author{Srimoyee Sen}%
\email{srimoyee08@gmail.com}
\affiliation{Department of Physics and Astronomy, Iowa State University, Ames, Iowa 50011, USA}%
\author{Lars Sivertsen}%
\email{lars@iastate.edu }
\affiliation{Department of Physics and Astronomy, Iowa State University, Ames, Iowa 50011, USA}%
\date{\today}

\maketitle

%%%%%%%%%% Merge with supplemental materials %%%%%%%%%%
%%%%%%%%%% Prefix a "S" to all equations, figures, tables and reset the counter %%%%%%%%%%
\renewcommand{\theequation}{S\arabic{equation}}
\renewcommand{\thefigure}{S\arabic{figure}}
\renewcommand{\thepage}{S\arabic{page}}
\renewcommand{\thesection}{S\arabic{section}}
\renewcommand{\thetable}{S\arabic{table}}
\newcommand{\hc}{\textrm{H.c.}}
\newcommand{\ii}{\textrm{i}}
\newcommand{\sigmavec}{\boldsymbol{\sigma}}
\newcommand{\n}{\mathbf{n}}
\makeatletter

\section{Quasienergy spectrum of $U_{\rm F}$}
In this section we will derive the quasienergy spectrum of the bulk Floquet Hamiltonian defined as 
\begin{equation}
    H_\text{F} = -\text{Im}[\log U_\text{F}] = -\text{Im}[\log(\exp(-\ii H_1 t_1)\exp(-\ii H_0 t_0))]\label{UF},
\end{equation}
where the SSH-Hamiltonians $H_i$ are defined as 
\begin{align}
\begin{split}
H_0&=2\sum^{N-1}_{j=0}(a_{2j}^{\dagger}a_{2j+1}+\hc)\\
H_1&=2\sum^{N-1}_{j=0}
(a_{2j+1}^{\dagger}a_{2j+2}+\hc).
\end{split}
\end{align}
Since we are considering the bulk spectrum, we have in the above and will in what follows assume periodic boundary conditions, $a_{2N}=a_0$. We define
\beq
a_{2j}=c_{2j},\,\, a_{2j+1}=d_{2j+1}.
\eeq
We proceed by performing a Fourier transform 
\begin{align}
    c_{2j} &= \frac{1}{\sqrt{N}}\sum_{k} e^{2\ii k j} c_k ,
    \nonumber
    \\
    d_{2j+1} &= \frac{1}{\sqrt{N}}\sum_{k} e^{+\ii k (2j+1)} d_k ,
\end{align}
where $0\leq k<\pi$.
The Hamiltonian written in momentum space are of the form
\begin{align}
H_0 &= 2\sum_k( c^\dagger_k d_k e^{\ii k}+\hc),
\nonumber
\\
H_1&= 2\sum_k (c^\dagger_k d_k e^{-\ii k}+\hc)
\end{align}
which can be rewritten in matrix form as 
\begin{align}
H_0 &= 2\begin{pmatrix}
    0 & e^{\ii k}
    \\
    e^{-\ii k} & 0
\end{pmatrix}
= 2\sigma_x\cos k-2\sigma_y\sin k\equiv 2 \sigmavec\cdot \n_0,
\nonumber
\\
H_1 &= 2\begin{pmatrix}
    0 & e^{-\ii k}
    \\
    e^{\ii k} & 0
\end{pmatrix}
= 2\sigma_x\cos k+2\sigma_y\sin k \equiv 2\sigmavec\cdot \n_1.
\end{align}
Here $\sigma_i$ are the Pauli spin matrices, and
\begin{equation}
    \n_0 = \begin{pmatrix}
    \cos k \\ -\sin k
    \end{pmatrix},
    \quad\quad\quad \n_1 =\begin{pmatrix}
        \cos k \\ \sin k
    \end{pmatrix},
    \quad\quad\quad 
    \sigmavec = 
    \begin{pmatrix}
        \sigma_x \\ \sigma_y
    \end{pmatrix}.
\end{equation}
Using the relation 
\begin{equation}
\exp(-\ii \theta(\sigmavec\cdot\n)) = \cos(\theta)-\ii(\sigmavec\cdot \n)\sin(\theta),
\end{equation}
we advance towards finding the eigenvalues of the Floquet operator by inserting the expressions for $H_0$ and $H_1$ into \eqref{UF}:
\begin{align}
U_{\text{F}} &= 
\exp(-\ii H_1 t_1)\exp(-\ii H_0 t_0)
\nonumber
\\
&=\exp(-2\ii t_1  (\sigmavec\cdot\n_1))\exp(-2\ii t_0  (\sigmavec\cdot \n_0)) 
\nonumber
\\
\end{align}
From here one can find the eigenvalues of $U_{\text{F}}$ %$\lambda$ of $U_\text{F}$ in the standard way by solving for zeros of the characteristic polynomial 
%\begin{equation}
%    \det(U_\text{F}-\lambda) = 0.
%\end{equation}
%Solving this equation and cleaning it up using trigonometric identities, we find
\begin{align}
\lambda =& \frac{1}{4}\Big[\cos(2 k -2 t_0-2 t_1)+2 \cos(2t_0-2t_1)-\cos(2k+2t_0-2t_1)
\nonumber
\\
&-\cos(2k - 2t_0 +2 t_1)+2\cos(2_0+2t_1)+\cos(2k +2t_0+2t_1)\Big]
\nonumber
\\
&\pm \ii \Big\{1-\frac{1}{16}[\cos(2 k -2 t_0-2 t_1)+2 \cos(2t_0-t_1)-\cos(2k+2t_0-2t_1)
\nonumber
\\
&\quad\quad\quad-\cos(2k - 2t_0 +2 t_1)+2\cos(2_0+2t_1)+\cos(2k +2t_0+2t_1)]^2\Big\}^{1/2}
\nonumber
\\
\equiv&e^{\pm \ii \theta}.
\end{align}
We can now obtain the eigenvalues of the Floquet Hamiltonian $H_F$ given by 
\begin{align}
\lambda_{H_\text{F}} =&\pm\theta 
\nonumber
\\
=& \pm\arccos\Big\{\frac{1}{4}\big[\cos(2 k -2 t_0-2 t_1)+2 \cos(2t_0-2t_1)-\cos(2k+2t_0-2t_1)
\nonumber
\\
&-\cos(2k - 2t_0 +2 t_1)+2\cos(2t_0+2t_1)+\cos(2k +2t_0+2t_1)\big]\Big\}.\label{theta}
\end{align}
In particular, note that for $t_0=\frac{\pi}{4}$ we have
\begin{align}
\lambda_{H_\text{F}} 
=&\pm\arccos\Big[-\sin(2t_1)\cos(2k)\Big].\label{theta t=0}
\end{align}
The arccos function has the feature
\begin{equation}
\arccos(-x) = \pi-\arccos(x),
\end{equation}
and therefore, for a fixed value of $t_1$, there is a $\pi$ pairing in the eigenvalues of $H_{\text{F}}$, i.e. if there is an eigenvalue $\lambda_{H_\text{F}}=\kappa$, there is another one at $\pi/T-\kappa$. This follows as $k$ takes values from $0$ to $\pi$. Furthermore, note that if we instead let $t_1=\pi/4$, we get the exact same result, except $t_0\to t_1$. Hence there is also $\pi$ pairing along the $t_1=\pi/4$ line. However, this pairing disappears with OBC, as the phases along this line have either zero or $\pi$ modes on the boundary, not both---the line $\frac{t_0}{T}=\frac{\pi}{4}$ is therefore the only region of the phase diagram with $\pi$-pairing independent of boundary conditions, so it is there that we expect the cleanest relationship between the Floquet and lattice-fermion models.

\section{Phases of $U_{\rm F}$}

In this section we discuss the phase diagram of the Floquet operator
\begin{align}
    U_{\rm F}=e^{-iH_1t_1}e^{-iH_0t_0},
\end{align}
where
\begin{align}
\begin{split}
H_0&=2\sum^{N-1}_{j=0}(a_{2j}^{\dagger}a_{2j+1}+\text{H.c.})\\
H_1&=2\sum^{N-2}_{j=0}
(a_{2j+1}^{\dagger}a_{2j+2}+\text{H.c.}).
\label{h}
\end{split}
\end{align}
Note that we have written the model with open boundary conditions to facilitate distinguishing the four phases by their edge modes.
In this section we will work in units such that $T=1$.
This model is directly analogous to the ``Class D" toy model of Ref.~\cite{vonKeyserlingk}, with the exception that the latter is composed of Majorana fermions instead of complex fermions.
We follow closely the analysis of Ref.~\cite{vonKeyserlingk}.
The strategy is to characterize $U_{\rm F}$ along the representative lines $t_0=0$, $t_1=0$, $t_0=\pi/2$, and $t_1=\pi/2$. Standard continuity arguments then imply that any point in parameter space that can be reached from these lines without closing the quasienergy gap must be in the same phase.

\subsection{Trivial and $0$ phases}

The trivial and $0$ phases are closely related to their undriven counterparts. The trivial phase is smoothly connected to the line $t_1=0$, where $U_{\rm F}$ reduces to evolution under the trivial  Hamiltonian $H_0$, which hybridizes all lattice sites in pairs. The $0$ phase is smoothly connected to the line $t_0=0$, where $U_{\rm F}$ reduces to evolution under the topological Hamiltonian $H_1$, which hybridizes bulk sites while leaving the edge sites $0$ and $2N-1$ decoupled.

\subsection{$0\pi$ phase}
To understand the $0\pi$ phase, it is useful to rewrite $H_0$ and $H_1$ via a Jordan-Wigner transformation as
\begin{align}
H_0&=\sum^{N-1}_{j=0}(X_{2j}X_{2j+1}+Y_{2j}Y_{2j+1})\\
H_1&=\sum^{N-2}_{j=0}
(X_{2j+1}X_{2j+2}+Y_{2j+1}Y_{2j+2}).
\label{hspin}
\end{align}
We then consider $U_{\rm F}$ along the line $t_1=\pi/2$:
\begin{align}
U_{\rm F} &= e^{-i\frac{\pi}{2}H_1}e^{-iH_0t_0}\\
&= \prod^{N-2}_{j=0} Z_{2j+1}Z_{2j+2}\ e^{-iH_0t_0}\\
&= U_{\rm edge}\, e^{-it_0(X_0X_1+Y_0Y_1+X_{2N-2}X_{2N-1}+Y_{2N-2}Y_{2N-1})}\, Q\,U_{\rm bulk}
\label{uFint1}
\end{align}
where $U_{\rm edge}=Z_0Z_{2N-1}$, $Q=\prod^{N-1}_{j=0}Z_{2j}Z_{2j+1}$, $U_{\rm bulk}=\prod^{N-2}_{j=1}e^{-it_0(X_{2j}X_{2j+1}+Y_{2j}Y_{2j+1})}$, and where we have used the identity
\begin{align}
\label{zzprod}
e^{-i\frac{\pi}{2}H_1} &= \prod^{N-2}_{j=0}e^{-i\frac{\pi}{2}(X_{2j+1}X_{2j+2}+Y_{2j+1}Y_{2j+2})}\\
&=\prod^{N-2}_{j=0}e^{-i\frac{\pi}{2}X_{2j+1}X_{2j+2}}e^{-i\frac{\pi}{2}Y_{2j+1}Y_{2j+2}}\\
&=-\prod^{N-2}_{j=0}X_{2j+1}X_{2j+2}Y_{2j+1}Y_{2j+2}\\
&=\prod^{N-2}_{j=0}Z_{2j+1}Z_{2j+2}.
\end{align}
Note that $Q$ constitutes a global symmetry as it commutes with $H_0$ and $H_1$.
Next, we transform Eq.~\eqref{uFint1} by the unitary operator
\begin{align}
W=e^{i\frac{t_0}{2}(X_0X_1+Y_0Y_1+X_{2N-2}X_{2N-1}+Y_{2N-2}Y_{2N-1})},
\end{align}
which commutes with $Q$, $U_{\rm edge}$, and $U_{\rm bulk}$, to obtain
\begin{align}
\tilde U_{\rm F}=W^\dagger U_{\rm F} W &= W^\dagger U_{\rm edge}\, e^{-it_0(X_0X_1+Y_0Y_1+X_{2N-2}X_{2N-1}+Y_{2N-2}Y_{2N-1})}W\, Q\, U_{\rm bulk}\\
&=U_{\rm edge}\, We^{-it_0(X_0X_1+Y_0Y_1+X_{2N-2}X_{2N-1}+Y_{2N-2}Y_{2N-1})}W\, U_{\rm bulk}\\
&=U_{\rm edge}\,Q\,U_{\rm bulk},
\end{align}
where we used that $U_{\rm edge}$ anticommutes with the generator of $W$. This demonstrates that the quasienergy spectrum of $U_{\rm F}$ factors into the product of the spectra of the commuting unitaries $U_{\rm edge}$, $Q$, and $U_{\rm bulk}$. Eigenstates can be labeled uniquely by their eigenvalues under the commuting quantities $U_{\rm edge}, Q, U_{\rm bulk}$, and under the additional bulk symmetry operator $P_{\rm bulk}=\prod^{N-2}_{j=1}X_{2j}X_{2j+1}$ and edge symmetry operator $P_{\rm edge}=X_0X_1X_{2N-2}X_{2N-1}$.

To see how $0$ and $\pi$ modes emerge, let us denote a reference eigenstate of $\tilde U_{\rm F}$ by $\ket{u_{\rm edge},p_{\rm edge}, p_{\rm bulk},q,E}$, where $E$ is an energy eigenvalue of $H_{\rm bulk}=\sum^{N-2}_{j=1}(X_{2j}X_{2j+1}+Y_{2j}Y_{2j+1})$ and where $u_{\rm edge},p_{\rm edge}, p_{\rm bulk},q=\pm 1$ are eigenvalues of  $U_{\rm edge},P_{\rm edge},P_{\rm bulk},$ and $Q$, respectively. (Note that all of these operators mutually commute with one another; we will therefore call them symmetries of $U_{\rm F}$.) Denote the quasienergy of this reference state by $\epsilon=i\ln(u_{\rm edge}\, q\, e^{-iEt_0})$. We can toggle between the four joint eigenstates of the edge operators $U_{\rm edge}$ and $P_{\rm edge}$ by acting with the operators $Z_0$, which anticommutes with $P_{\rm edge}$ and commutes with all other symmetries, and $X_0X_1$, which anticommutes with $U_{\rm edge}$ and commutes with all other symmetries. This generates the following ``$0\pi$-multiplet" of states:
\begin{center}
\begin{tabular}{|c|c|}
\hline
   State  & Quasienergy \\
   \hline\hline
    $\ket{u_{\rm edge},p_{\rm edge}, p_{\rm bulk},q,E}$ & $\epsilon$\\
    \hline
    $Z_0\ket{u_{\rm edge},p_{\rm edge}, p_{\rm bulk},q,E}=\ket{u_{\rm edge},-p_{\rm edge}, p_{\rm bulk},q,E}$ & $\epsilon$\\
    \hline
    $X_0X_1\ket{u_{\rm edge},p_{\rm edge}, p_{\rm bulk},q,E}=\ket{-u_{\rm edge},p_{\rm edge}, p_{\rm bulk},q,E}$ & $\epsilon+\pi$\\
    \hline
    $Y_0X_1\ket{u_{\rm edge},p_{\rm edge}, p_{\rm bulk},q,E}=\ket{-u_{\rm edge},-p_{\rm edge}, p_{\rm bulk},q,E}$ & $\epsilon+\pi$\\
\hline
\end{tabular}
\end{center}
We refer to $Z_0$ as a zero-mode operator since it commutes with $U_{\rm F}$, and $X_0X_1$ as a $\pi$-mode operator because it anticommutes with $U_{\rm F}$. 
Note that there is another zero- and $\pi$-mode operator at the other end of the chain: $Z_{2N-1}=U_{\rm edge}Z_0$ and $X_{2N-2}X_{2N-1}=P_{\rm edge}X_0X_1$. Thus we could have alternatively chosen to label states in this multiplet by, say, eigenvalues of $Z_0$ and $Z_{2N-1}$ instead of $U_{\rm edge}$ and $P_{\rm edge}$, and to toggle between these eigenvalues with $X_0X_1$ and $X_{2N-2}X_{2N-1}$.
%Note that this phase is protected by the bulk symmetry $P$. Without enforcing this symmetry, one can write down local operators that mix different topological sectors. For example, the two states at quasienergy $\epsilon$ can hybridize via the operator $Z_1$, which anticommutes with $P$ but commutes with the other symmetries.

\subsection{$\pi$ phase}

The $\pi$ phase can be treated similarly to the $0\pi$ phase. Here we consider the representative point $t_0=\pi/2$, where
\begin{align}
U_{\rm F} &= e^{-iH_1t_1}e^{-i\frac{\pi}{2}H_0}\\
&=e^{-iH_1t_1}\, Q\\
&=e^{-iH_1t_1}\left(\prod^{N-2}_{j=0}Z_{2j+1}Z_{2j+2}\right)Z_0Z_{2N-1}\\
&=e^{-iH_1\left(t_1+\frac{\pi}{2}\right)}Z_0Z_{2N-1}\\
&=U_{\rm bulk}\, U_{\rm edge},
\end{align}
where we have used Eq.~\eqref{zzprod}. Note that the resulting model also has the global symmetry $P=\prod^{2N-1}_{j=0}X_{2j}X_{2j+1}$ with eigenvalues $p=\pm 1$. We can repeat the analysis of the $0\pi$ phase to demonstrate the existence of $\pi$ modes. To do this, let $\ket{u_{\rm edge},p,E}$ be a reference eigenstate labeled by the symmetry eigenvalues $u_{\rm edge},p=\pm 1$ and an energy eigenvalue $E$ of $H_1$. The associated quasienergy is $\epsilon=i\ln(u_{\rm edge}\, e^{-iEt_1})$. The eigenvalues of $U_{\rm edge}$ can be toggled by the $\pi$-mode operator $X_0$, leading to the following ``$\pi$-doublet":
\begin{center}
\begin{tabular}{|c|c|}
\hline
   State  & Quasienergy \\
   \hline\hline
    $\ket{u_{\rm edge},p,E}$ & $\epsilon$\\
    \hline
    $X_0\ket{u_{\rm edge},p,E}=\ket{-u_{\rm edge},p,E}$ & $\epsilon+\pi$\\
    \hline
\end{tabular}
\end{center}

\subsection{Mapping to two copies of the Majorana model}

A simpler way to verify that the phase diagram of $U_{\rm F}$ matches that of the Majorana model of Ref.~\cite{vonKeyserlingk} is to observe that the former amounts to two copies of the latter. To see this we write the complex fermions $a_{j}$ in terms of Majorana fermions $\alpha_j,\beta_j$ in the standard way:
\begin{align}
    a_j = \frac{1}{2}(\alpha_j+i\beta_j).
\end{align}
Substituting this into Eq.~\eqref{h} gives
\begin{align}
\begin{split}
H_0&=\sum^{N-1}_{j=0}(i\, \alpha_{2j}\beta_{2j+1}-i\, \beta_{2j}\alpha_{2j+1})\\
H_1&=\sum^{N-2}_{j=0}
(i\, \alpha_{2j+1}\beta_{2j+2}-i\, \beta_{2j+1}\alpha_{2j+2}).
\end{split}
\end{align}
This is precisely two copies of the aforementioned Majorana model, with one of the copies time-reversed with respect to the other. We note that the two copies are not strictly independent, since performing a U(1) transformation on the complex fermions leads to a nontrivial mixing of the Majorana operators:
\begin{align}
a_j \to e^{i\theta}a_j = (\cos\theta\, \alpha_j+\sin\theta\, \beta_j)+i(\sin\theta\, \alpha_j -\cos\theta\, \beta_j).
\end{align}
This mixing is necessary in order for the two copies of the Majorana model to manifest the $U(1)$ symmetry of the complex-fermion model.

\section{Domain wall in $\eta$ in the Wilson-Dirac model}
Here we consider a spatially dependent Wilson-Dirac Hamiltonian with a domain wall in the parameter $\eta=\frac{t_1}{T}-\frac{\pi}{4}$ that tunes between the trivial and $0\pi$ phases. To implement this we pick the following domain wall configuration with $+$ and $-$ subscripts denoting the regions $x_1>0$ and $<0$
\beq
m_{\pm}&=&-\sin(2\eta_{\pm})\nonumber\\
R_{\pm}&=&\frac{1+\sin(2\eta_{\pm})}{2}\nonumber\\
\eta_+&=&-\eta_->0. 
\eeq
We can now look for edge modes on the domain wall by solving the equation of motion
\beq
(-iR\gamma_1\nabla_1+m-\frac{R}{2}\nabla_1^2)\psi=0
\eeq
with $\gamma_0=\sigma_2$ and $\gamma_1=-i\sigma_1$ and $m$ and $R$ are spatially dependent, i.e. taking the above specified values for $x_1>0$ and $<0$. We can solve this EOM with an ansatz of 
\beq
\psi=\begin{pmatrix}1\\
1
\end{pmatrix}\varphi
\eeq
which reduces the EOM to
\beq
&&-\frac{\varphi(x_1+1)-\varphi(x_1-1)}{2}+\frac{m}{R}\varphi-\frac{\varphi(x_1+1)+\varphi(x_1-1)-2\varphi(x_1)}{2}=0\nonumber\\
&\implies&\varphi(x_1+1)=\left(1+\frac{m}{R}\right)\varphi.
\eeq
So, we find normalizable solutions for all of the parameter space of interest, i.e. $\frac{\pi}{4}>\eta>0$ given by
\beq
\varphi(x_1)&=&\left(1+\frac{m}{R}\right)^{x_1}\nonumber\\
&=&\begin{cases}
\left(1+\frac{m_+}{R_+}\right)^{x_1}\,\,\,\,\,\,\, \text{for $x_1>0$}\nonumber\\
\left(1+\frac{m_-}{R_-}\right)^{x_1}\,\,\,\,\,\,\, \text{for $x_1<0$}.
\end{cases}
\eeq
This corresponds to a localized mode with localization length
\begin{align}
\xi_\pm = \frac{-1}{\ln\left(1-\frac{|m_{\pm}|}{R_{\pm}}\right)}
\end{align}
to either side of the domain wall.

\bibliography{floquet.bib}